\documentclass[12pt,onecolumn]{IEEEtran} 
\def\BibTeX{{\rm B\kern-.05em{\sc i\kern-.025em b}\kern-.08em
     T\kern-.1667em\lower.7ex\hbox{E}\kern-.125emX}}

\usepackage{cite,amsmath,amscd,amssymb,graphicx,latexsym,multicol}
\usepackage{algorithm,algorithmic}
\usepackage{comment}

\newtheorem{proposition}{Proposition}
\newenvironment{thmproof}[1]
{\noindent\hspace{2em}{\it #1 }}
{\hspace*{\fill}~\QED\par\endtrivlist\unskip}

\centerfigcaptionstrue

\newcommand{\Exp}{\mathbb E}
\newcommand{\expect}[1]{{\Exp}\left\{#1\right\}}
\newcommand{\expqcnd}[2]{{\Exp_q}\left\{\left. #1 \,\right|\, #2\right\}}

\newcommand{\sm}[1]{\left\langle #1 \right\rangle}
\newcommand{\expcnd}[2]{{\Exp}\left\{\left. #1 \,\right|\, #2\right\}}
\newcommand{\mse}{{ \mathcal{E} }} 
\newcommand{\vrc}{{ \mathcal{V} }} 
\newcommand{\snr}{ \Gamma }
\newcommand{\qXYS}{q_{\X|\Y,\S}}
\newcommand{\X}{\mathbf{X}}
\newcommand{\Y}{\mathbf{Y}}
\renewcommand{\S}{\mathbf{S}}
\newcommand{\x}{\mathbf{x}}
\newcommand{\y}{\mathbf{y}}
\newcommand{\intd}{{\,\normalfont{\text d}}} 
\newcommand{\expbr}[1]{ \exp \left[ {#1} \right] } 

\begin{document}
\bibliographystyle{ieeetr}
%


\title{A Unified Approach to Energy-Efficient Power Control in Large CDMA Systems\thanks{This research was
supported by the National Science Foundation under Grants
ANI-03-38807 and CCF-0644344, as well as by DARPA under Grant
W911NF-07-1-0028. Parts of this work have been presented at the
$2{nd}$ International Workshop on Signal Processing for Wireless
Communications (SPWC'04), King's College London, in June 2004, and
at the $43{rd}$ Annual Allerton Conference on Communication,
Control and Computing, University of Illinois at Urbana-Champaign,
in September 2005.}}

\author{Farhad Meshkati, Dongning Guo, H. Vincent Poor, and Stuart C.
Schwartz\vspace{-0.8cm}\thanks{ {{F. Meshkati was with the
Department of Electrical Engineering, Princeton University. He is
currently with QUALCOMM Inc., 5775 Morehouse Dr., San Diego, CA
92121 USA (e-mail: meshkati@qualcomm.com).}}}\thanks{ {{D. Guo is
with the Department of Electrical Engineering and Computer
Science, Northwestern University, Evanston, IL 60208 USA (e-mail:
{dguo@northwestern.edu}). }}} \thanks{{{H. V. Poor and S. C.
Schwartz are with the Department of Electrical Engineering,
Princeton University, Princeton, NJ 08544 USA (e-mail:
{\{poor,stuart\}@princeton.edu}).}}}  }

\maketitle

\begin{abstract}
  A unified approach to energy-efficient power control is proposed for
  code-division multiple access (CDMA) networks. The approach is applicable
  to a large family of multiuser receivers including the matched
  filter, the decorrelator, the linear minimum mean-square error
  (MMSE) receiver, and the (nonlinear) optimal detectors. It exploits the linear relationship
that has been shown to exist between the transmit power and the
output signal-to-interference-plus-noise ratio (SIR) in the
large-system limit. It is shown that, for this family of
receivers, when users seek to selfishly maximize their own energy
efficiency, the Nash equilibrium is SIR-balanced. In addition, a
unified power control (UPC) algorithm for reaching the Nash
equilibrium is proposed. The algorithm adjusts the user's transmit
powers by iteratively computing the large-system multiuser
efficiency, which is independent of instantaneous spreading
sequences. The convergence of the algorithm is proved for the
matched filter, the decorrelator, and the MMSE receiver, and is
demonstrated by means of simulation for an optimal detector.
Moreover, the performance of the algorithm in finite-size systems
is studied and compared with that of a conventional power control
scheme, in which user powers depend on the instantaneous spreading
sequences.
\end{abstract}

\begin{keywords}
  Code-division multiple access (CDMA), energy efficiency, game
  theory, large systems, multiuser detection, multiuser
  efficiency, Nash equilibrium, power control.

\end{keywords}

\section{Introduction}

Power control is used for interference management and resource
allocation in both the downlink and the uplink of
code-division multiple access (CDMA) networks
 \cite{Foschini93, Bambos95, Yates95, Ulukus97,
HanlyTsePC, ShitzVerdu01, Andrews02, Leung04}.
In particular,
in the uplink 
each user 
transmits just enough
power to achieve the required quality of service (QoS) without
causing 
excessive interference in the network.
In recent years, power
control for multiuser receivers has attracted much attention
due to their
superior performance as compared to the
single-user matched filter. 
For example, power control algorithms for the linear minimum
mean-square error (MMSE) receiver and successive interference
cancellation receivers have been proposed in \cite{Ulukus97} and
\cite{Andrews02}, respectively. In the proposed schemes, the
output signal-to-interference-plus-noise ratio (SIR) of each is
measured and then the user's transmit power is
 adjusted to achieve the desired 
SIR.

Power control is 
conventionally 
modeled as an 
optimization problem under some quality of service (QoS) constraints.
A practically appealing scheme is to minimize
the total transmit power 
under the constraint that the 
output SIR of each user is above some lower bound.
Under reasonable assumptions it is shown that
the total transmit power is minimized
when the SIR requirements are met with equality \cite{Foschini93}. 
Another approach
is to choose the transmit powers 
in such a way as to
maximize the spectral efficiency (in bits/s/Hz) of the network,
where the optimal 
strategy is essentially a
water-filling scheme 
\cite{ShitzVerdu01}. 

In recent years, game
theory has been used to study power control in CDMA networks. 
In particular, power control can be modeled as a non-cooperative
game in which each user selfishly maximizes its own utility. The
strategy chosen by one user affects the performance of other users
in the network through multiple-access interference
 (e.g.,
\cite{GoodmanMandayam00,Saraydar02,Xiao01,Alpcan,Sung,Meshkati_Tcomm,
MeshkatiJSAC}). 


Reference \cite{Meshkati_Tcomm}
studies the cross-layer design problem of joint power
control and multiuser detection using game theory.
The utility function in this
case measures the number of bits transmitted per joule of energy
consumed, which 
is particularly suitable for
energy-constrained networks. Focusing on linear receivers, the
Nash equilibrium for the 
non-cooperative game has been derived. A Nash
equilibrium refers to a set of strategies for which no user can
unilaterally improve its own utility \cite{FudenbergTiroleBook91}.
It is also shown in \cite{Meshkati_Tcomm} that for all
linear receivers, the transmit powers of the
users are SIR-balanced at Nash equilibrium
(i.e., all users have the same output SIR).

This work extends the results in \cite{Meshkati_Tcomm} to a much
larger family of receivers (including nonlinear ones) in the
so-called {\em large-system} regime, where the number of users and
the spreading factor are large with a given ratio. This is due to
results in \cite{Guo05} where a linear relationship between the
input power and the output SIR is shown to exist for a family of
multiuser detectors in the large-system limit. Members of this
family include well-known receivers such as the matched filter
(MF), the decorrelator (DEC), the linear MMSE receiver, as well as
the individually optimal (IO) and jointly optimal (JO) multiuser
detectors.\footnote{The individually optimal detector minimizes
the error probability of detecting an individual symbol whereas
the jointly optimal detector minimizes the probability of error
for detecting the entire vector of all users' symbols at a symbol
interval \cite{VerduBook98}.  The jointly optimal detector is
often referred to as the maximum likelihood (ML) detector.} By
exploiting the linear relationship, which is characterized by the
multiuser efficiency, we propose a unified power control (UPC)
algorithm for reaching the Nash equilibrium.
The convergence of the
proposed algorithm is proved for linear receivers and is
demonstrated by means of simulation for an optimal 
detector.

Since the UPC algorithm is
based on large-system results, it does not depend on the
instantaneous spreading sequences.
The true SIR, however, does depend on the spreading sequences.
Therefore, the SIR achieved by the UPC algorithm deviates from the
target SIR as the spreading sequence changes from one
symbol to the next. 
 This in turn will result in a loss in the
utility.  The performance of the UPC algorithm in finite-size
systems is studied and it is shown that if the spreading factor is
reasonably large, the SIR achieved by the UPC algorithm stays
close to the target SIR most of the time which means that the loss
is insignificant. 

The rest of the paper is organized as follows.
Section~\ref{multiuser} provides the system model and relevant
results in multiuser detection.
Section~\ref{gametheory} discusses the game-theoretic approach to
power control.
The UPC 
algorithm for reaching Nash equilibrium
is proposed
and 
studied
 in Section~\ref{power control}.
 The performance of the UPC algorithm in
finite-size systems is studied in Section~\ref{performance}.
Simulation results are presented in Section~\ref{simulation}
before  conclusions are drawn in Section~\ref{conclusions}.

\section{Multiuser Detection and Power Control} \label{multiuser}

\subsection{System Model and Multiuser Detection}

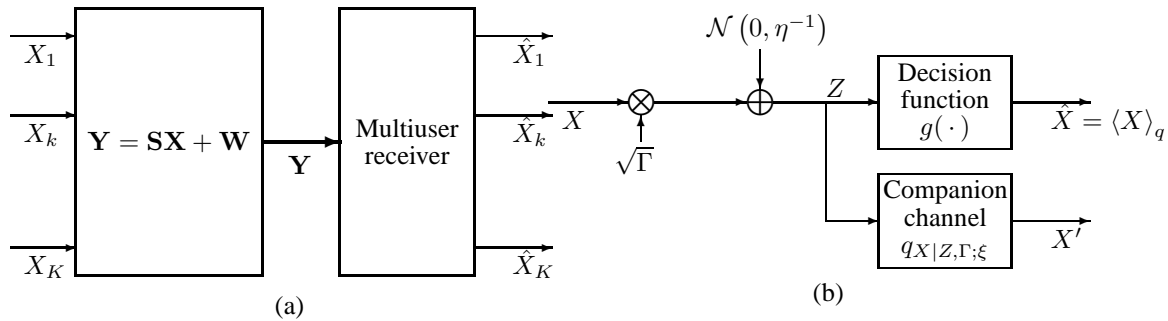
\begin{figure}
  \small{
  \begin{center}
    \begin{picture}(200,110)(60,-5)
      \thinlines
      \put(60,60){\vector(1,0){25}}
      \put(65,50){$X_k$}
      \put(60,10){\vector(1,0){25}}
      \put(60,90){\vector(1,0){25}}
      \put(65,0){$X_K$}
      \put(65,80){$X_1$}
      \thicklines
      \put(85,0){\framebox(70,100){\parbox{70pt}{
            \begin{center}
              $\mathbf{Y} = \mathbf{S}\mathbf{X}+ \mathbf{W}$
            \end{center}}}}
      \put(155,50){\vector(1,0){30}}
      \put(165,38){$\mathbf{Y}$}
      \thicklines
      \put(185,0){\framebox(50,100){\parbox{50pt}{
            \begin{center}Multiuser\vspace{-0.4cm}\\receiver\end{center}}}}
      \thinlines
      \put(235,60){\vector(1,0){30}}
      \put(250,50){$\hat{X}_k$}
      \put(235,10){\vector(1,0){30}}
      \put(235,90){\vector(1,0){30}}
      \put(250,0){$\hat{X}_K$}
      \put(250,80){$\hat{X}_1$}
      \put(160,-15){(a)}
    \end{picture}
    \begin{picture}(200,130)(60,-40)
      \put(60,30){\vector(1,0){30}}
      \put(65,20){$X$}
      \put(95,30){\makebox(0,0){$\bigotimes$}}
      \put(85,4){$\sqrt{\Gamma}$}
      \put(95,15){\vector(0,1){10}}
      \put(100,30){\vector(1,0){35}}
      \put(140,30){\makebox(0,0){$\bigoplus$}}
      \put(140,50){\vector(0,-1){15}}
      \put(120,55){$\mathcal{N}\left(0,\eta^{-1}\right)$}
      \put(145,30){\vector(1,0){40}}
      \put(165,32){$Z$}
      \put(185,12.5){\framebox(50,35){\parbox{50pt}{
            \begin{center}Decision\vspace{-0.4cm}\\function\vspace{-0.4cm}\\ $g(\,\cdot\,)$\end{center}}}}
      \put(185,-32.5){\framebox(50,35){\parbox{50pt}{
            \begin{center}Companion\vspace{-0.4cm}\\channel\vspace{-0.4cm}\\ $q_{X|Z,\Gamma;\xi}$\end{center}}}}
      \put(235,30){\vector(1,0){30}}
      \put(250,20){$\hat{X}=\sm{X}_q$}
      \put(165,-15){\vector(1,0){20}}
      \put(165,-15){\line(0,1){45}}
      \put(235,-15){\vector(1,0){30}}
      \put(250,-25){$X'$}
      \put(160,-45){(b)}
    \end{picture}
  \end{center}
  }
  \caption{(a) The DS-CDMA system with multiuser detection, where only
    the output for user $k$ is shown.  (b) The equivalent single-user
    Gaussian channel for user $k$; also shown is the companion
    channel.}
  \label{f:mud}
\end{figure}

Consider the uplink of a synchronous DS-CDMA system with $K$
users and spreading factor $N$.
Let $p_k$, $h_k$, and $\sigma^2$
represent the transmit power, channel gain and the background
noise variance (including other-cell interference)
respectively, for user $k$.  
The received signal-to-noise ratio (SNR) for user $k$ is then
\begin{equation}\label{eq1}
  \Gamma_k = \frac{p_k h_k}{ \sigma^2}.
\end{equation}
The received signal (after chip-matched
filtering) sampled at the chip rate over one symbol duration can
be represented as
\begin{equation} \label{eq2}
  {\mathbf{Y}} =
  \S\X + \mathbf{W}
  = \sum^K_{k=1} \sqrt{\Gamma_k} X_k {\mathbf{s}}_k +
  {\mathbf{W}}  , 
\end{equation}
where $\mathbf{s}_k$ and $X_k$ are the spreading sequence and
transmitted symbol of user $k$, respectively, and
 $\mathbf{W}\sim \mathcal{N} (0,\mathbf{I})$ is
the noise vector consisting of independent standard Gaussian
entries (see Fig.~\ref{f:mud}(a)). Random spreading sequence is
assumed for all
users, and the input symbols $X_k$ 
are assumed to be independent and identically
distributed (i.i.d.) with unit variance.

As is shown in Fig.~\ref{f:mud}(a), the multiuser receiver
produces an estimate $\hat{X}_k$ for each user $k$.
Suppose the receiver is a linear filter $\mathbf{c}_k$, the detection
output can be represented as a sum of three independent components:
\begin{align}
  \hat{X}_k
  &= \mathbf{c}_k^T \mathbf{Y} \\
  &= \sqrt{\Gamma_k} X_k + \mathsf{MAI}_k + V_k
  \label{e:zl}
\end{align}
i.e., the desired signal, the MAI and Gaussian background noise.  The
quality of the output is effectively measured in terms of the output
SIR,
\begin{equation}  \label{eq:gk}
  \gamma_k = \frac{p_k h_k ({\mathbf{c}}_k^T {\mathbf{s}}_k)^2}
  {\sigma^2 {\mathbf{c}}_k^T {\mathbf{c}}_k
    + \sum_{j\neq k} p_j h_j ({\mathbf{c}}_k^T{\mathbf{s}}_j)^2}
  = \eta_k \Gamma_k
\end{equation}
because the multiple-access interference ($\mathsf{MAI}_k$) is
asymptotically Gaussian (e.g., \cite{GuoVer02IT}).  The
degradation in SNR due to the MAI is known as the {\em multiuser
efficiency}, denoted by $\eta_k$.  In fact, \eqref{e:zl} can be
regarded as an equivalent single-user Gaussian channel for user
$k$ as depicted in Fig.~\ref{f:mud}(b), in which the function
$g(\cdot)$ is an identity mapping, e.g., $g(z)=z$, $\forall z$.

\subsection{Power Control}

We model power control  as a non-cooperative game in which
each user tries to maximize its own utility (see e.g.,
\cite{GoodmanMandayam00,Saraydar02,Xiao01,Alpcan,Sung,Meshkati_Tcomm,
MeshkatiJSAC}).
We follow \cite{GoodmanMandayam00} to define
the utility of user $k$ as 
\begin{equation}\label{eq10}
    u_k = \frac{T_k}{p_k} \;\, \text{bits/joule},
\end{equation}
where $T_k$ is the throughput of user $k$, i.e.,
the net number of information bits
delivered correctly per unit time (sometimes referred to as
\emph{goodput}).
This utility function captures 
the
tradeoff between throughput and battery life and is particularly
suitable for applications where saving power is more important
than achieving a high throughput.

The throughput for user $k$ can be quantified as 
\begin{equation}\label{eq11}
    T_k = \frac{L}{M} R f(\gamma_k)  ,
\end{equation}
where $L$ and $M$ are the number of information bits and the total
number of bits in a packet, respectively (i.e., $L-M$ bits of
overhead); $R$ is the transmission rate and $f(\gamma_k)$ is the
efficiency function representing the packet success rate. 
The underlying 
assumption is that if a packet is in error, 
it
will be retransmitted. The efficiency function, $f(\gamma_k)$, is
assumed to be increasing and S-shaped\footnote{An increasing
function is S-shaped if there is a point above which the function
is strictly concave, and below which the function is strictly
convex.} (sigmoidal) with $f(\infty)=1$. We also require that
$f(0)=0$ to ensure that $u_k=0$ when $p_k=0$. These assumptions
are valid in many practical systems (see \cite{Meshkati_Tcomm} for
further details). Combining \eqref{eq10} and \eqref{eq11}, user
$k$'s utility is given by
\begin{equation}\label{eq11b}
  u_k = \frac{L \, R\, f(\gamma_k)}{ M\, p_k}  \ .
\end{equation}
Using a sigmoidal efficiency function, it can be shown that the
utility function in \eqref{eq11b} is quasiconcave.\footnote{A
function is quasiconcave if there exists a point below which the
function is non-decreasing, and above which the function is
non-increasing.}

%

Suppose each user is allowed to vary its transmit power only in order
to selfishly maximize the utility, the power control game is described
as
\begin{equation}\label{eq11b2}
    \max_{p_k} u_k \ \ \ \textrm{for} \ \ k=1,\cdots,K.
\end{equation}
For this non-cooperative game, a Nash equilibrium is a set of
transmit powers $(p_1,\dots,p_K)$
for which no user can unilaterally improve its own
utility (i.e., 
 a stable state). It has been
shown in \cite{Meshkati_Tcomm} that the
following proposition holds
for all linear receivers.  
\begin{proposition} \label{prop1}
The utility-maximizing strategy for user $k$  is given by
${p_k^*=\min(\hat{p}_k, P_{max})}$, where $\hat{p}_k$ is the
transmit power that results in an output SIR equal to $\gamma^*$,
which is the solution to $f(\gamma)=\gamma f'(\gamma)$, and
$P_{max}$ is the maximum allowed power. Furthermore, the proposed
power control game has a unique Nash equilibrium.
\end{proposition}

This proposition implies that, at Nash equilibrium, the users'
transmit powers are SIR-balanced.
The key point in proving Proposition~\ref{prop1} is
that for all linear receivers there is a linear relationship
between the output SIR and transmit power of a user. If we take
the derivative of the utility function in \eqref{eq10} with
respect to the transmit power and equate it to zero, we have
\begin{equation}\label{eq13}
    p_k  \frac{\partial \gamma_k}{\partial p_k}  f'(\gamma_k) - f(\gamma_k)
    =0 .
\end{equation}
Based on the expression of the output SIR \eqref{eq:gk},
we can write
\begin{equation}\label{eq14}
  \frac{\partial \gamma_k}{\partial p_k} =\frac{\gamma_k}{p_k} \ .
\end{equation}
Therefore, the utility of user $k$ is maximized when
$\gamma_k=\gamma^*$, the (positive) solution to
\begin{equation}\label{eq15}
f(\gamma)=\gamma f'(\gamma) \ .
\end{equation}
The existence of a Nash equilibrium is guaranteed because of the
quasiconcavity of the utility function and its uniqueness is
because of the one-to-one relationship between the user's transmit
power and output SIR.

\section{Large Multiuser Systems and Power Control}\label{gametheory}

This paper studies power control in {\em large} systems.
Mathematically, we consider the so-called {\em large-system limit},
where both the number of users and the spreading factor tend to
infinity but with their ratio
converging to a positive number, i.e., $K/N\rightarrow\alpha$.
Since, in a large system, the dependence between the SNRs
$\Gamma_k$ is reasonably weak, we assumed that $\Gamma_k$ are
i.i.d.\ with distribution $P_\Gamma$ at a given time.  Moreover,
$P_\Gamma$ varies slowly over time as the timescale of power
control is much larger than the symbol interval.

In general, the multiuser efficiency depends on the received SNRs, the
spreading sequences as well as the type of detector.  However, in the
asymptotic case of large systems, the dependence on the spreading
sequences vanishes and the received SNRs affect $\eta$ only through
their distribution.
In particular,
the multiuser efficiency of the
matched filter and the decorrelator are obtained as
\begin{align}
  \eta^{mf} &= \frac{1}{1+\alpha {\mathbb{E}}\{\Gamma\}} \label{eq6}\\
  \eta^{dec} &= 1-\alpha  \ \ \ \textrm{for} \ \ \alpha<1 \label{eq7}
\end{align}
while the efficiency of the (linear) MMSE receiver is the unique
solution to the following fixed-point equation
\begin{equation}
  \frac{1}{\eta^{mmse}} = 1 +\alpha {\mathbb{E}}\left\{
    \frac{\Gamma}{1+\eta^{mmse}\Gamma}\right\} \label{eq8}
\end{equation}
where the expectation is over $P_\Gamma$.

The analysis of the SIR using \eqref{eq:gk} does not directly apply to
nonlinear receivers because the output of such receivers cannot be
decomposed as a sum of the desired signal and independent
interference; neither is the output asymptotically Gaussian.
Remarkably,
reference \cite{Guo05}
finds that, under mild assumptions, the output of a nonlinear receiver
converges in the large-system limit to a simple monotone function of a
``hidden'' Gaussian statistic conditioned on the input, i.e.,
\begin{equation} \label{e:zf}
  \hat{X}_k \rightarrow g\big( \sqrt{\Gamma_k}\,X_k + U_k \big)
\end{equation}
where $U_k \sim \mathcal{N} \big(0,\eta^{-1}\big)$ is independent of
$X_k$.
By applying an inverse of this function to the detection output
$\hat{X}_k$, an equivalent conditionally Gaussian statistic $Z_k =
\sqrt{\Gamma_k}\, X_k + U_k$ is recovered.  Each symbol $X_k$
traverses an equivalent single-user Gaussian channel, so that the
output SIR (defined for the equivalent Gaussian statistic $Z_k$)
completely characterizes the system performance.  This result is
referred to as the ``decoupling principle.''  The equivalent
channel is illustrated in Fig.~\ref{f:mud}(b).  Indeed, as far as
the posterior probability $P_{X_k|\hat{X}_k}$ is concerned, the
multiuser model (Fig.~\ref{f:mud}(a)) and the single-user model
(Fig.~\ref{f:mud}(b)) are asymptotically indistinguishable.

\subsection{Posterior Mean Estimators}

The decoupling principle is shown in \cite{Guo05} to hold for a broad
family of multiuser receivers, called the {\em posterior mean
  estimators} (PME).  Given the observation $\Y$ and the spreading
matrix $\S$, a PME computes the mean value of
some 
posterior probability distribution $\qXYS$, which is conveniently
denoted as
\begin{equation}  \label{e:vxq}
  \sm{\X}_q = \expqcnd{ \X }{ \Y, \S }
\end{equation}
where $\Exp_q\{\cdot\}$ stands for the expectation with respect to the
measure $q$.

In this work, the posterior $q_{\X|\Y,\S}$ supplied to the PME is
induced from the following postulated CDMA system,
\begin{equation} \label{e:pch}
  \Y = \S\X' + \varrho\mathbf{W}
\end{equation}
which differs from the actual channel \eqref{eq2} by only the input
and the noise variance.  In particular, the components of $\X'$ are
i.i.d.\ with distribution $q_X$, and the postulated noise level
$\varrho$ serves as a control parameter.  The posterior $q_{\X|\Y,\S}$
is determined by $q_X$ and $q_{\Y|\X,\S}$ according to Bayes' formula
\begin{equation}  \label{e:vxqq}
  \qXYS(\x|\y,\S) = \frac{ q_{\X}(\x) q_{\Y|\X,\S}(\y|\x,\S) }
  { \int q_{\X}(\x) q_{\Y|\X,\S}(\y|\x,\S) \,\intd \x }.
\end{equation}

Indeed, the PME so defined is parameterized by $(q_X, \varrho)$ and can
be regarded as the optimal detector for the postulated multiuser
system \eqref{e:pch}.  In case the postulated posterior $q_{\X|\Y,\S}$
is identical to $p_{\X|\Y,\S}$, the PME is a soft version of the
individually optimal detector.  The postulated posterior, however, can
also be chosen such that the PME becomes one of many other detectors,
including but not limited to the matched filter, decorrelator, linear
MMSE receiver, as well as the optimal
detectors.  Thus the concept of PME is generic and versatile.

\subsection{Decoupling Principle for PME}
\label{s:gmd}

Let $q_{Z|X,\snr;\xi}$ represent the input--output relationship of a
scalar Gaussian channel,
\begin{equation}      \label{e:qzx}
    q_{Z|X,\snr;\xi}(z|x,\snr;\xi) = \sqrt{\frac{\xi}{2\pi}}
      \expbr{ -\frac{\xi}{2} \left(z - \sqrt{\snr} \,x\right)^2 }.
\end{equation}
Similar to that in the multiuser setting, by postulating the input
distribution to be $q_X$, a posterior probability distribution
$q_{X|Z,\snr;\xi}$ is induced from $q_X$ and $q_{Z|X,\snr;\xi}$
using Bayes' formula (cf.~\eqref{e:vxqq}).  Thus we have a
single-user companion channel defined by $q_{X|Z,\snr;\xi}$, which
outputs a random variable $X'$ given the channel output $Z$
(Fig.~\ref{f:mud}(b)).  A (generalized) single-user PME is defined
naturally as:
\begin{equation}
  \sm{X}_q
  = \expqcnd{ X }{ Z,\snr;\xi }
\label{e:xq}
\end{equation}
The probability law of the composite system depicted by Fig.~\ref{f:mud}(b)
 is determined by $\snr$, 
$\eta$ and
$\xi$.

\begin{proposition}[\cite{Guo05}]  \label{th:dp}
  Fix $(\alpha, P_\snr, p_X, q_X, \varrho)$.  The joint probability
  distribution of $(X_k, \sm{X_k}_q)$ converges to the joint
  probability distribution of $\big(X, g\big(\sqrt{\Gamma_k}\,X+U\big)
  \big)$ where $U\sim \mathcal{N} \big(0, \eta^{-1} \big)$ and
  \begin{equation}    \label{eq:f}
    g(z) = \expqcnd{ X }{ Z=z,\snr;\xi },
  \end{equation}
  where the multiuser efficiency $\eta$ satisfies together with $\xi$ the
  coupled equations:
  \begin{subequations}    \label{e:ex}%
    \begin{eqnarray}
      \eta^{-1} &=& 1\; + \alpha \, \expect{ \snr\cdot\mse(\snr;\eta,\xi) },
      \label{e:e} \\
      \xi^{-1} &=& \!\varrho^2 + \alpha\,\expect{\snr\cdot\vrc(\snr;\eta,\xi) },
      \label{e:x}
    \end{eqnarray}
  \end{subequations}
  where the expectations are taken over $P_\snr$.  Here we define the
  mean squared error of the PME as
\begin{equation}  \label{e:gmse}
  \mse(\snr;\eta,\xi) =
  \expcnd{ \left( X - \sm{X}_q \right)^2 }{ \snr;\eta,\xi },
\end{equation}
and also define the variance of the companion channel as
\begin{equation}  \label{e:vrc}
  \vrc(\snr;\eta,\xi) =
  \expcnd{ \left( X' - \sm{X}_q \right)^2 }{ \snr;\eta,\xi }.
\end{equation}
  In case of multiple
  solutions to \eqref{e:ex}, $(\eta,\xi)$ is chosen to minimize the
  free energy (see \cite{Guo05}).
\end{proposition}

Proposition \ref{th:dp} reveals that, from an individual user's
viewpoint, the input--output relationship of the multiuser channel
concatenated with the multiuser receiver is asymptotically identical
to that of the scalar Gaussian channel with a (nonlinear) decision
function.  The key performance measure is the equivalent SIR $\eta$ of
the scalar channel, which is easy to compute by solving the
fixed-point equation \eqref{e:ex}, since the functions $\mathcal{E}$
and $\mathcal{V}$ can be easily computed given the distribution of the
transmitted symbols and the type of receiver; and the expectations are
taken with respect to the received SNR distribution.

By choosing appropriate parameters $(q_X,\rho)$, the PME can be made
to represent the matched filter, the decorrelator, and the linear MMSE
detector as well as the individually and jointly optimal detectors.
The multiuser efficiencies of the linear receivers are given as
\eqref{eq6}--\eqref{eq8}.
The multiuser efficiency of
the individually optimal detector is found to satisfy the fixed-point
equation
\begin{equation}
  \frac{1}{\eta^{io}} = 1 +\alpha {\mathbb{E}}\left\{ \Gamma - \Gamma
    \!\! \int_{-\infty}^{+\infty} \frac{e^{-\frac{z^2}{2}}}{\sqrt{2\pi}}
    \tanh \left(\eta^{io}\Gamma - z
      \sqrt{\eta^{io}\Gamma}\right)\textrm{d}z \right\}.
  \label{eq9}
\end{equation}

This result generalizes to multirate systems as well
\cite{GuoMultirate}.

In principle, the multiuser efficiency may be different for
different users with different types of receivers. As the uplink
is the focus of this work, it is assumed that all users use the
same type of receiver and hance are subject to the same multiuser
efficiency.

\subsection{Power Control}

The decoupling principle implies that in large systems there is a
linear relationship between the transmit power and the output SIR
of each user, i.e.,
\begin{equation}\label{eq3}
  \gamma_k = \eta \Gamma_k \ ,
\end{equation}
where the multiuser efficiency $\eta$ 
depends on the SNR distribution $P_\Gamma$ rather
than the individual $\Gamma_k$'s. This is mainly due to the fact
that in a large system, as one user's transmit power varies, 
 the
interference seen by the user essentially stays the same as long
as the overall distribution of the received powers remains the
same.

Even though (\ref{eq3}) is a large-system result,
it is a very good approximation for most finite-size
systems of practical interest.
The linear relationship \eqref{eq3} implies that, for
the family of PME receivers, \eqref{eq14} is satisfied in large
systems. Therefore, the argument given above for the linear
receivers extends to a larger family of receivers that includes
some nonlinear receivers such as the individually and joint
optimal multiuser detectors.

This means that, in the asymptotic
case of large systems, the Nash equilibrium for the power control
game in \eqref{eq11b} is SIR-balanced for all the receivers
belonging to the family of PME for which \eqref{eq3} is satisfied.
In other words, the Nash equilibrium is reached when each user
transmits at a power level that achieves an output SIR equal to
$\gamma^*$, the solution of $f(\gamma)=\gamma f'(\gamma)$.
It is interesting to note that the
Nash equilibrium SIR, $\gamma^*$, is independent of the type of
receiver and depends only on physical-layer parameters such as
modulation, coding and packet size.

Clearly, 
any deviation of a user's output SIR from $\gamma^*$ results in a
loss in the user's utility, or equivalently, its energy
efficiency. For example, fixing other users' transmit powers, if
the output SIR of  user $k$ is equal to $\hat{\gamma}$
(instead of $\gamma^*$), the resulting utility 
is
given by
\begin{equation}\label{eq15c}
    \hat{u}_k = \left( \frac{ \gamma^*
    f(\hat{\gamma})}{\hat{\gamma} f(\gamma^*)}\right) u_k^*  \
    ,
\end{equation}
where $u_k^*$ is the maximum utility for user $k$ (corresponding
to $\gamma_k=\gamma^*$).

\section{Unified Power Control Algorithm}\label{power control}

In this section, we propose a unified power control algorithm for
reaching the Nash equilibrium of the power control game, which is
applicable to the family of PME receivers.
The algorithm
iteratively adjusts the transmit powers in order to reach an
output SIR equal to $\gamma^*$.
While we assume balanced SIRs here as a property of the Nash
equilibrium,
the UPC algorithm is
also applicable to the case of unequal
target SIRs.


The UPC Algorithm carries out the following iteration:
\begin{enumerate}
\item Let $n=0$, start with initial powers $p_1(0),\cdots,p_K(0)$.

\item Use \eqref{e:ex} and the power profile to compute the multiuser
  efficiency, $\eta(n)$.

\item For user $k$, update the powers according to
  \begin{equation} \label{e:pk}
    p_k(n+1) = \frac{1}{\eta(n)}
    \frac{\gamma^* \, \sigma^2}{ h_k}.
  \end{equation}


\item \emph{n}=\emph{n}+1, stop if convergence; otherwise, go to Step
  2.
\end{enumerate}

In Step 2, while finding an analytical expression for
the multiuser efficiency is
difficult for most multiuser detectors, 
it can be easily
obtained from \eqref{e:ex} using numerical methods.
Note that (\ref{e:ex})
need to be solved only once per iteration for each
user. The uplink receiver (e.g., base station) can, for example,
compute $\eta$ and feed it back to the user terminal.
The above algorithm is applicable to
a large family of receivers which includes many popular receivers
such as the matched filter, the decorrelator, and the MMSE
detector as well as the individually and jointly optimal multiuser
detectors. Note, for example, that for a matched filter receiver,
the UPC algorithm becomes the same as the bilinear power control
algorithm proposed in \cite{Gajic04} for minimizing the SIR error.
In fact, one may also extend Proposition \ref{th:dp} to the case
that different users use different type of receivers and Algorithm
1 will apply by replacing $\eta$ by individual efficiencies $\eta_k$.

It is apparently a dilemma that the success of the power control
scheme depends on the assumption that the SNR distribution is
fixed, while the purpose of power control is to adjust
the powers which may affect the SNR distribution.  This,
however, can be resolved naturally in practice, where the number
of users is finite, by an iterative power control algorithm and by
replacing the expectations in \eqref{e:ex} with
an average over all users' received SNRs (or their estimates).  For
example, \eqref{eq8} can be expressed as
\begin{equation}
\frac{1}{\eta} = 1 +
\frac{\alpha}{K} \sum^K_{k=1} \frac{\Gamma_k}{1+\eta \Gamma_k}
\end{equation}
At any rate, the UPC scheme provides a viable mechanism to
approach the Nash equilibrium.

There are certainly other practical
issues that need further investigation. For example, the UPC
algorithm requires the background noise and other-cell
interference power to be measured. This could be done by silencing
the users in a cell in a coordinated manner. In addition, while we
will show in Section~\ref{simulation} that the UPC algorithm
converges very quickly, the convergence rate and effects of
estimation/measurement errors on the convergence of the algorithm
require further analysis.

We now prove the convergence of the UPC algorithm for the matched
filter, the decorrelator, and the MMSE receiver.\footnote{The
convergence analysis for a general receiver remains an open problem
because \eqref{e:ex} is difficult to
work with.}  Let
${\mathbf{\Gamma}}=\left[\Gamma_1, \cdots , \Gamma_K\right]$ and
define an interference function,
\begin{equation}\label{eq16}
  I({\mathbf{\Gamma}})=\frac{\gamma^*}{\eta({\mathbf{\Gamma}})}\ .
\end{equation}
The dependence of the multiuser efficiency on ${\mathbf{\Gamma}}$
is explicitly shown here.
By \eqref{e:pk} and \eqref{eq16}, the UPC algorithm can be
expressed as
\begin{equation}\label{eq17}
  {\Gamma}_k(n+1)=I\left({\mathbf{\Gamma}}(n)\right),
  \quad k=1,\dots,K.
\end{equation}
\begin{proposition} \label{prop2}
For the matched filter, the decorrelator, and the MMSE receiver,
if there exists a $\hat{{\mathbf{\Gamma}}}$ such that
$\hat{\Gamma}_k \geq
I(\hat{{\mathbf{\Gamma}}})$, $k=1,\dots,K$,
 then for every initial
vector ${\mathbf{\Gamma}}(0)$, the recursion \eqref{eq17}
converges to the unique fixed point solution of
$\Gamma^*_k=I({\mathbf{\Gamma^*}})$, $k=1,\dots,K$.
Furthermore, for any feasible $\hat{{\mathbf{\Gamma}}}$ (i.e.,
$\hat{\Gamma}_k \geq I(\hat{\mathbf{\Gamma}})$ for all $k$),
$\Gamma^*_k \leq \hat{\Gamma}_k$ for all $k$.
\end{proposition}
\begin{thmproof}{Proof:}
The existence of a $\hat{{\mathbf{\Gamma}}}$
implies that a feasible SNR
vector exists for achieving $\gamma^*$.
It suffices then to show that
$I({\mathbf{\Gamma}})$ is a standard interference function
\cite{Yates95},  i.e., for all ${\mathbf{\Gamma}}$ with $\Gamma_k
\geq 0$ for all $k$, the following three properties are satisfied:
1) Positivity: $I({\mathbf{\Gamma}})>0$; \
2) Monotonicity: If ${\Gamma'}_k\geq \Gamma_k$ for all $k$, then
$I({\mathbf{\Gamma'}})\geq I({\mathbf{\Gamma}})$;
\ 3) Scalability: For all $\theta>1$,
$\theta I({\mathbf{\Gamma}})>I({\mathbf{\theta\Gamma}})$.
Evidently, it is equivalent to showing the three properties for
${\hat{I}({\mathbf{\Gamma}})=1/{\eta ({\mathbf{\Gamma}})}}$.

Positivity of $\hat{I}({\mathbf{\Gamma}})$ is trivial by \eqref{e:e}
for all receivers since $\eta \in [0,1]$.

Consider first the matched filter.  The multiuser efficiency is given by
\eqref{eq6},
where
$\mathbb{E}\{\Gamma\} = \sum^K_{k=1} \Gamma_k / K$.  If
${\mathbf{\Gamma'}}\geq{\mathbf{\Gamma}}$, then
${\mathbb{E}}\{\Gamma'\}\geq{\mathbb{E}}\{\Gamma\}$ and hence
$\hat{I}({\mathbf{\Gamma'}})\geq \hat{I}({\mathbf{\Gamma}})$. To prove the
third property, note that, for $\theta>1$, $\hat{I}(\theta
{\mathbf{\Gamma}})=1+\alpha {\mathbb{E}}\{\theta \Gamma\} < \theta
+\alpha \theta {\mathbb{E}}\{\Gamma\}= \theta
\hat{I}({\mathbf{\Gamma}})$.

Consider now the decorrelator, the multiuser efficiency of which is
constant $\eta= 1-\alpha > 0$ for $\alpha<1$.
Properties 2 and 3 are trivial.


For the MMSE receiver, the multiuser efficiency is the solution to
\eqref{eq8}, or equivalently, the unique solution of
\begin{equation}  \label{eq:et}
  \eta +\alpha {\mathbb{E}}\left\{ \frac{1}{\frac{1}{\eta\Gamma}+1}\right\}=1.
\end{equation}
Note that the left-hand side of \eqref{eq:et}
increases if both $\eta$ and $\Gamma$
increase. Thus if ${\mathbf{\Gamma'}}\geq{\mathbf{\Gamma}}$, we
must have $\eta({\mathbf{\Gamma'}})\leq \eta({\mathbf{\Gamma}})$
to maintain the equality. Hence,~$\hat{I}({\mathbf{\Gamma'}})\geq
\hat{I}({\mathbf{\Gamma}})$. To prove the third property, let us define
$\eta'=\eta({\mathbf{\theta \Gamma}})$ and $\eta''=\theta \eta'$,
where $\theta>1$. Therefore,
\begin{equation}
\eta'' + \alpha \theta
{\mathbb{E}}\left\{ \frac{1}{\frac{1}{\eta''
\Gamma}+1}\right\}=\theta.
\end{equation}
Showing ${\theta
\hat{I}({\mathbf{\Gamma}}) > \hat{I}(\theta {\mathbf{\Gamma}})}$
is equivalent to showing $\eta'' > \eta$. Since
 the left-hand side of \eqref{eq:et}
is increasing in $\eta$, and
\begin{equation}
{\eta'' +\alpha
{\mathbb{E}}\left\{ \frac{1}{ \frac{1}{ \eta'' \Gamma}+1}\right\}
=1 + \left( 1-\frac{1}{\theta} \right) \eta'' > 1},
\end{equation}
we  must have $\eta''>\eta$. Therefore,
${\theta \hat{I}({\mathbf{\Gamma}}) > \hat{I}(\theta {\mathbf{\Gamma}})}$.
This completes the proof.
\end{thmproof}

\section{Performance Evaluation and Discussion} \label{performance}

The UPC algorithm described in Section \ref{power control} is
a large-system approach and hence independent of the spreading
sequences. Therefore, after convergence, the transmit powers need
not be updated as long as the channel gains remain static.
Evidently, the actual SIRs depend on the spreading sequences (see
for example \eqref{eq:gk}).
Consequently, as the spreading sequence changes, the output SIRs
achieved by the UPC algorithm fluctuate around the target SIR.
This fluctuation results in a loss in energy efficiency (see
\eqref{eq15c}).

The question of interest is: if
the UPC algorithm is used, how close will the SIRs be to the
target SIR?
In the following, we study the performance of the UPC
algorithm for finite-size systems and compare it with that of an
SIR-based algorithm.
We focus on the decorrelator
and the linear MMSE receiver.

\subsection{The Decorrelator}


For the decorrelator, it is sensible to assume $\alpha<1$.  The
large-system multiuser efficiency is given by $\eta^{dec}=1-\alpha$.
Hence
the SNR dictated by the UPC algorithm is
\begin{equation}\label{eq22}
    \Gamma_k^* = \Gamma_{dec}^*
= \frac{\gamma^*}{1-\alpha} , \quad 
    k=1,\cdots,K,
\end{equation}
which should lead to an SIR of $\gamma^*$ in the large-system limit.

However, the actual output SIR for a finite-size system is given by
\begin{equation}\label{eq23}
  \gamma_k= \left(\frac{\gamma^* }{1-\alpha} \right)
  \bigg/ { \left[\big(\tilde{\mathbf{S}}^T
    \tilde{\mathbf{S}}\big)^{-1} \right]_{kk} },
\end{equation}
where $\tilde{\mathbf{S}}=\left[{\mathbf{s}}_1 , {\mathbf{s}}_2 ,\cdots,
{\mathbf{s}}_K\right]$
and $[\,\cdot\,]_{kk}$ extracts the $k$th diagonal entry of a matrix.


It has been shown that in large systems, 
the
distribution of ${1} / {
\left[({\mathbf{S}}^T{\mathbf{S}})^{-1} \right]_{kk}}$ can be
approximated by a beta distribution with parameters $(N-K+1, K-1)$
\cite{Muller}. As a result, 
the probability
density function 
of $\gamma_k$ is given approximately by
\begin{equation}\label{eq24}
    f_{\gamma_{dec}}(z) =
    \left(\frac{1}{\Gamma_{dec}^*}\right)^{N-1} \frac{
    z^{N-K}(\Gamma_{dec}^*-z)^{K-2}}{B(N-K+1, K-1)} \ \
\end{equation}
where $z\leq \Gamma_{dec}^*$ and
 $B(a,b)=\int_0^1 t^{a-1} (1-t)^{b-1} \textrm{d}t$.
Therefore, as the
spreading sequences change from symbol to symbol, the probability
that $\gamma_k$ stays within $\Delta$~dB of $\gamma^*$ is given by
\begin{align}\label{eq25}
  P_{\Delta, dec}
  &= \textrm{Pr}\left\{ |\gamma_{dec} \textrm{(dB)} -
    \gamma^* \textrm{(dB)} | \leq \Delta\right\} \\
  &= \int_{\gamma_L}^{\gamma_H} f_{\gamma_{dec}}(z)\textrm{d}z \ ,
\end{align}
where $\gamma_L=10^{-\frac{\Delta}{10}}\gamma^*$ and
$\gamma_H=10^{\frac{\Delta}{10}}\gamma^*$.

Alternatively, the fluctuation of the actual 
SIR around $\gamma^*$ can be approximated less accurately by a
Gaussian distribution with variance \cite{TseZeitouni}
\begin{equation}
\zeta^2 = \frac{2{\gamma^*}^2
  \alpha}{(1-\alpha)N} \ ,
\end{equation}
i.e., $\gamma_{dec}\sim{\mathcal{N}}\big(\gamma^*, \zeta^2\big)$.
Therefore, the probability that $\gamma_k$ stays within
$\Delta$~dB of $\gamma^*$ is approximately given by
\begin{equation}\label{eq25c}
    P^{\textrm{norm}}_{\Delta, dec} =
     \Phi\left(\frac{\gamma_H-\gamma^*}{\zeta}\right)-
   \Phi\left(\frac{\gamma_L-\gamma^*}{\zeta}\right) ,
\end{equation}
where $\Phi(\cdot)$ is the cumulative distribution function of the
standard Gaussian distribution.

\subsection{The MMSE Receiver}

If the linear MMSE receiver is used and all users  have the same
target SIR, $\gamma^*$, the steady-state SNRs will be identical to
$\Gamma^*=\gamma^*/\eta$ after the UPC algorithm converges,
where the multiuser efficiency is given by
\begin{equation}\label{eq26}
  \eta=  \frac{1-\alpha}{2}-\frac{1}{2\Gamma}
  + \frac{1}{2} \sqrt{\left(1-\alpha\right)^2 +
    \frac{2(1+\alpha)}{\Gamma} +\frac{1}{\Gamma^2}}    .
\end{equation}
It can be shown that 
the fluctuation of the true SIR around $\gamma^*$
is approximately Gaussian with variance 
\cite{TseZeitouni, KimHonig}: 
\begin{equation}
  \zeta^2 = \frac{1}{N} \,
  \frac{2{\gamma^*}^2}{1-\alpha\big(\frac{\gamma^*}{1+\gamma^*}\big)^2} \ .
\end{equation}
The probability that $\gamma_k$ stays within $\Delta$~dB of $\gamma^*$
admits a similar expression to \eqref{eq25c} using the function
$\Phi(\cdot)$.

It is seen from these approximations that the variance of
fluctuations of SIR decreases as $1/N$. In the following section,
we demonstrate the performance of the UPC algorithm using
simulations and also investigate the accuracy of the theoretical
approximations discussed above.

\section{Numerical Results}\label{simulation}

Consider the uplink of a randomly spread
DS-CDMA system with $K$ users and spreading factor $N$.
The 
noise variance is assumed to be
$\sigma^2 = {1.6\times 10^{-14}}$. 
We use $f(\gamma)=(1-e^{-\gamma})^M$ as the
efficiency function\footnote{This is a useful example for the
efficiency function and serves as an approximation to the packet
success rate that is very reasonable for moderate to large packet
sizes.} 
with $\gamma^*=6.4$ (=8.1 dB).
\begin{figure}
\begin{center}
\includegraphics[width=5in]{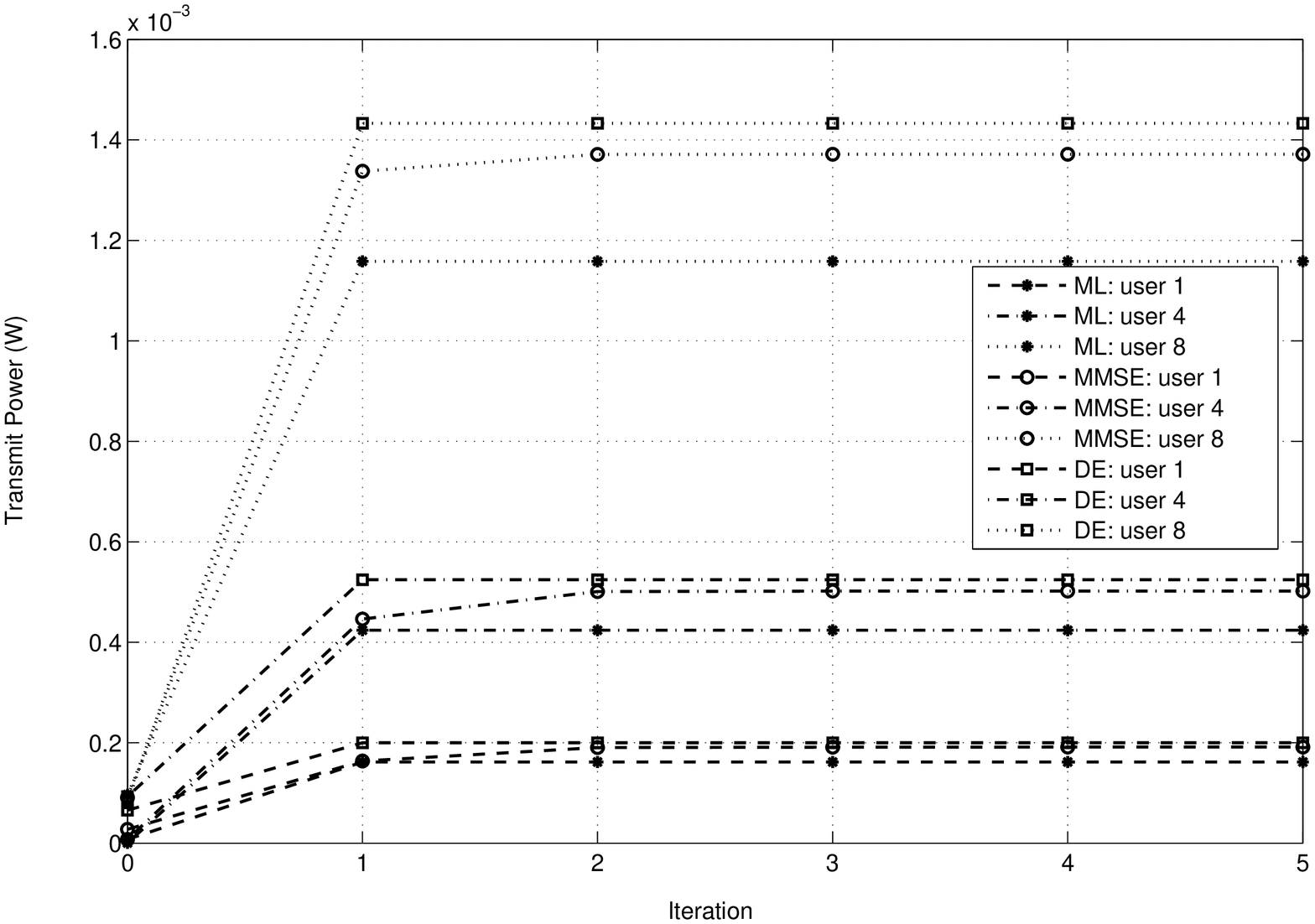}
\end{center}
\caption{Transmit powers for the ML, MMSE, and
decorrelator, using the UPC algorithm ($N=32$ and
$K=8$).} \label{figsim1}
\end{figure}

We first demonstrate the convergence of the UPC algorithm
assuming
$K=8$ and $N=32$. The channel gain for user $k$ is given by
$h_k={0.1}\times{d_k^{-4}}$ where $d_k$ is the distance of user $k$
from the uplink receiver (e.g., base station).
Assume $d_k=100+10k$ in meters. We implement the UPC algorithm
for the decorrelator and the MMSE receiver as well as the maximum
likelihood detector.

Fig.~\ref{figsim1} plots the transmit powers for users 1, 4, and 8
at the end of each iteration. It is seen that for all three
receiver types, the UPC algorithm converges quickly to
steady-state values. The results are similar when the initial
power values and/or $K$ and $N$ are changed. It is also observed
that the steady-state transmit powers for the decorrelator and the
MMSE receiver are close to those of the ML detector (in this case,
the difference is less than 22\%). This means that in terms of
energy efficiency, which is quantified by the utility achieved at
Nash equilibrium, the decorrelator and the MMSE receiver are
almost as good as the ML detector.

\begin{figure}
\begin{center}
\includegraphics[width=5in]{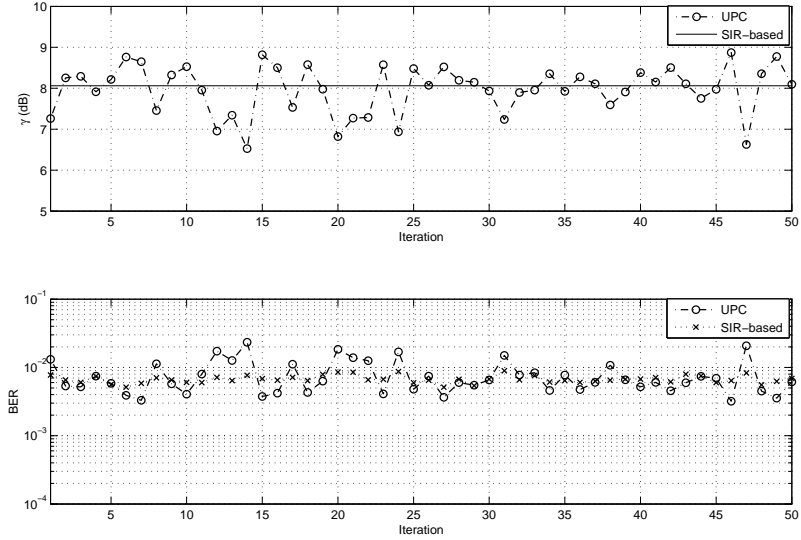}
\end{center}
\caption{ User 1 output SIR and BER for the UPC algorithm and
SIR-based algorithm with the MMSE receiver ($N=32$ and $K=8$).}
\label{MMSE-gamma-ber}
\end{figure}
%

We next investigate the fluctuation of the SIR and bit-error-rate
(BER) achieved by the (large-system) UPC algorithm against perfect
power control where the SIR is computed using instantaneous spreading
sequences (labeled SIR-based in plots).
Fig.~\ref{MMSE-gamma-ber} shows the SIR and bit-error-rate (BER)
of user~1  
using the MMSE
detector. It is seen that the SIR-based algorithm achieves the
target SIR, $\gamma^*$, at all time whereas the output SIR for the
UPC algorithm fluctuates around the target SIR as the spreading
sequences change. Also, the fluctuations in the BER are larger
when the UPC algorithm is used. 

\begin{figure}
\begin{center}
\includegraphics[width=5in]{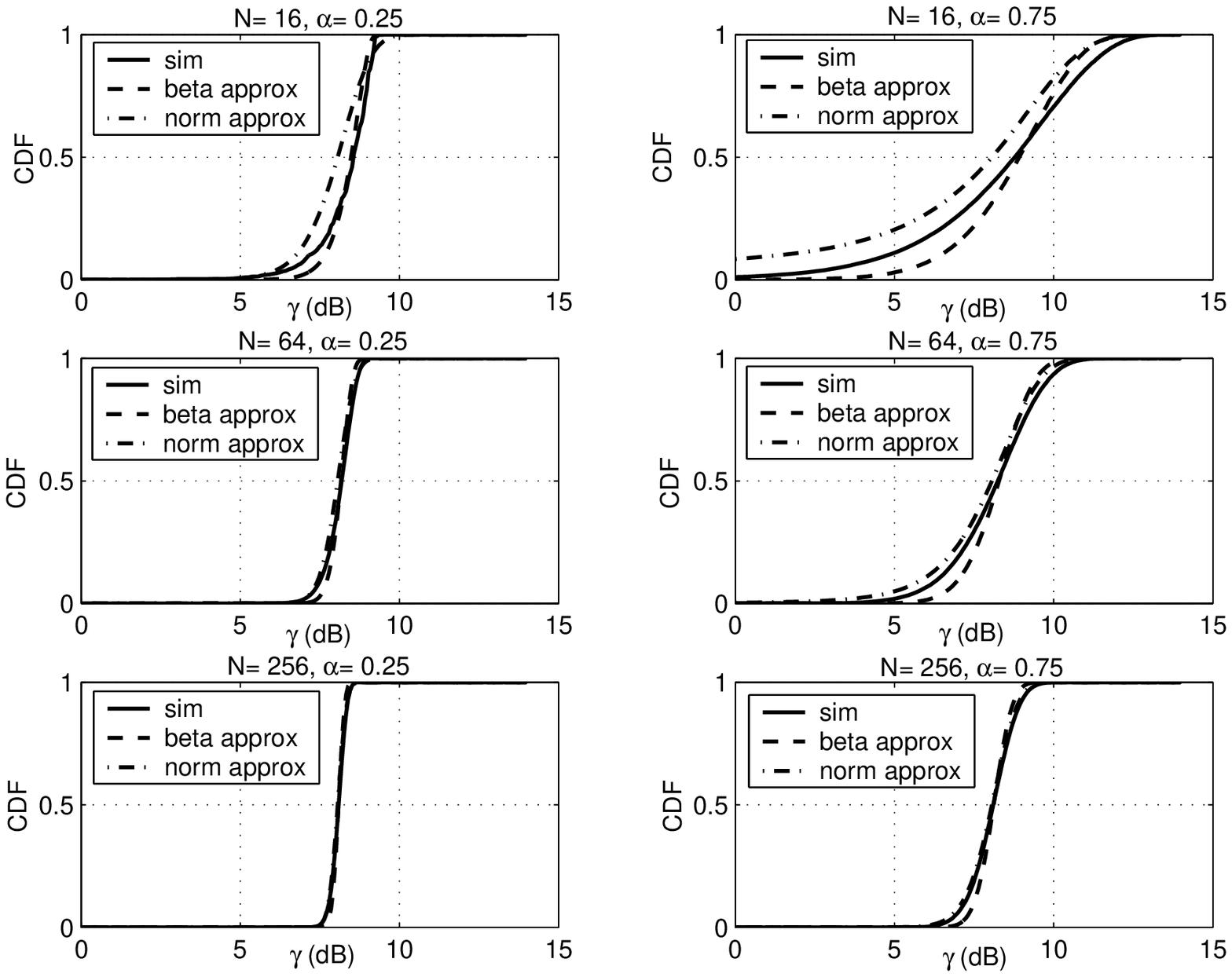}
\end{center}
\caption{CDFs of $\gamma$ for the decorrelator.} \label{figsim4de}
\end{figure}

\begin{figure}
\begin{center}
\includegraphics[width=5in]{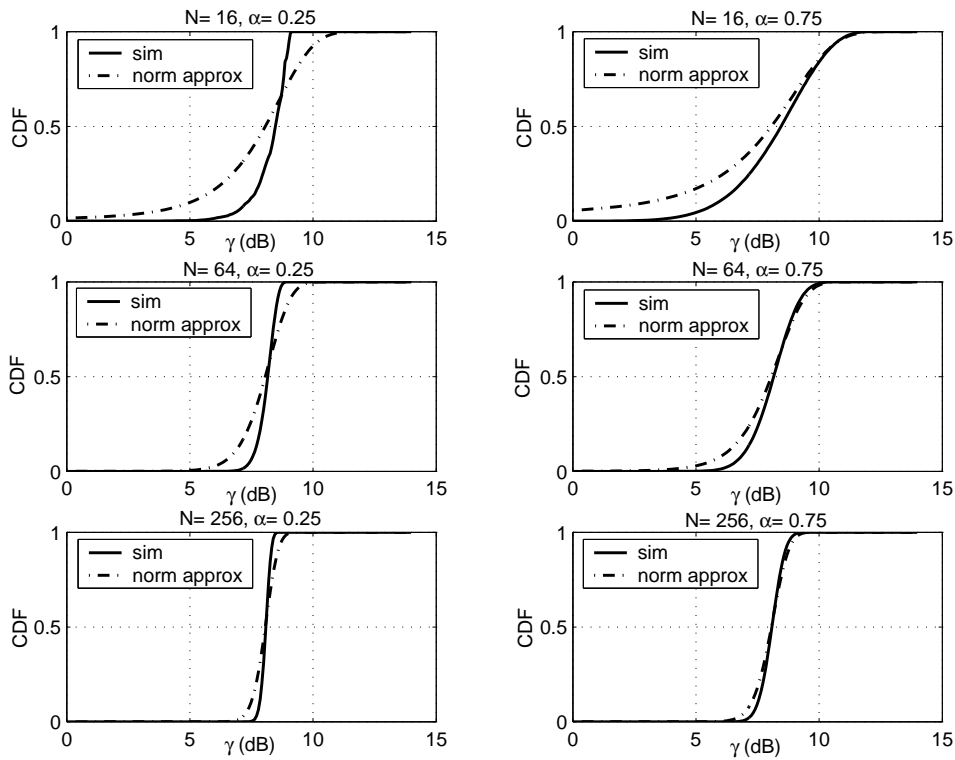}
\end{center}
\caption{CDFs of $\gamma$ for the MMSE receiver.} \label{figsim4mmse}
\end{figure}

To evaluate the accuracy of the theoretical approximations given
in Section~\ref{performance},
Figs.~\ref{figsim4de}~and~\ref{figsim4mmse}
  plot the cumulative
distribution function (CDF) of the output SIR $\gamma$ for the
decorrelator and the MMSE receiver with different spreading factors
and both low and high system loads.
The plots show CDFs obtained from simulation
(based on 100,000 realizations) as well as those predicted by the
theoretical approximations given in Section~\ref{performance}. It
is seen from the figures that the theoretical approximations
become more accurate as the spreading factor increases. Also,
the approximations are generally more accurate when the system load
is low.
Note that, for the decorrelator, the
approximation based on a beta distribution is slightly more
accurate than the one based on a Gaussian distribution, especially
for small spreading factors and large system loads.

\begin{table*}
 {\footnotesize {\begin{center} \caption{
Summary of results for the decorrelator and the MMSE
receiver}\label{table1}
\begin{tabular}{|c |c |c |c |c |c |c |c |c|c|c|}
  \hline
   N    & $P_{\textrm{1dB}, dec}^{sim} $  & $P_{\textrm{1dB}, dec}^{beta}$  & $P_{\textrm{1dB}, dec}^{norm}$ & $P_{\textrm{1dB},
   dec}^{sim} $ & $P_{\textrm{1dB}, dec}^{beta}$ & $P_{\textrm{1dB}, dec}^{norm}$  & $P_{\textrm{1dB}, MMSE}^{sim}$
   & $P_{\textrm{1dB}, MMSE}^{norm} $ & $P_{\textrm{1dB},
   MMSE}^{sim} $ & $P_{\textrm{1dB}, MMSE}^{norm}$  \\
    & $\alpha=0.25$ & $\alpha=0.25$  & $\alpha=0.25$ & $\alpha=0.75$ &
    $\alpha=0.75$ & $\alpha=0.75$ & $\alpha=0.25$ & $\alpha=0.25$ & $\alpha=0.75$ &
    $\alpha=0.75$\\
  \hline \hline
   16 & 0.77 & 0.87 & 0.74 & 0.28 & 0.19 & 0.30  & 0.93 & 0.46 & 0.41 & 0.33 \\
   64 & 0.98 & 1.0 &  0.97 & 0.54 & 0.64  & 0.55 & 0.99 & 0.76 & 0.74 & 0.61 \\
   256 & 1.0 & 1.0 &  1.0  & 0.87 & 0.96  & 0.87 & 1.0  & 0.98 & 0.98 & 0.91 \\
   \hline
\end{tabular}
\end{center}}}
\end{table*}

To quantify the discrepancies between the simulation results and
the theoretical approximations, we compute $P_{\Delta, dec}$
and $P_{\Delta, MMSE}$ using the CDFs obtained from simulations as
well as those predicted by theory (see \eqref{eq25} and \eqref{eq25c}).
Table \ref{table1} shows the results for
different spreading factors and system loads for $\Delta=1$~dB. The
numbers in the table represent the probability that $\gamma$ is
within 1~dB of $\gamma^*$. Based on \eqref{eq15c}, a 1-dB increase
in the output SIR results in 10\% loss in the user's utility. The
probabilities obtained by simulation suggest that the UPC
algorithm performs better for the MMSE receiver than for the
decorrelator. It is also seen from the table that when the
spreading factor is small, the fluctuations in the output SIR are
considerable, especially when the system load in high. The
performance improves as the spreading factor increases. For
example, for the MMSE receiver, when $N=256$ and $\alpha=0.75$,
the SIR stays within 1~dB of the target SIR $98\%$ of the time. It
is also observed that the theoretical approximations for the MMSE
detector are pessimistic. For the decorrelator, while
approximating the SIR by a beta distribution is more accurate (see
Fig. \ref{figsim4de}), the values obtained for $P_{\Delta,MMSE}$
by the Gaussian approximation are closer to the simulation results.
This is because the slope of the CDF of $\gamma$ is closer to the
slope of the Gaussian CDF. Since $P_{\Delta,MMSE}$
heavily depends on the slope of the CDF, it is more accurately
predicted by the Gaussian approximation (rather than the beta
approximation).


\section{Conclusion} \label{conclusions}

A unified approach to energy-efficient power control in large
systems is proposed, which is applicable to a large family of
linear and nonlinear multiuser receivers. The approach exploits
the linear relationship between the transmit power and the output
SIR in large systems. Taking a non-cooperative game-theoretic
approach with emphasis on energy efficiency, it is shown that the
Nash equilibrium is SIR-balanced not only for linear receivers but
also for some nonlinear receivers such as the individually and
jointly optimal multiuser detectors. In addition, a unified power
control algorithm for reaching the Nash equilibrium is proposed.
It would be straightforward to extend the unified approach to
multirate and multicarrier systems based on related large-system
results \cite{GuoMultirate, GuoMulticarrier}.

\end{document}